\documentstyle[12pt,epsfig]{article}
\newcommand{\scaption}[1]{\caption{\protect{\footnotesize  #1}}}

\newcommand{\average}[1]{\mbox{$ \langle #1 \rangle $}}
\newcommand{\av}[1]{\mbox{$ \langle #1 \rangle $}}
\newcommand{\kjet}{\mbox{$k_{T\rm{jet}}$}}
\newcommand{\xjet}{\mbox{$x_{\rm{jet}}$}}
\newcommand{\Ejet}{\mbox{$E_{\rm{jet}}$}}
\newcommand{\thjet}{\mbox{$\theta_{\rm{jet}}$}}
\newcommand{\pjet}{\mbox{$p_{T\rm{jet}}$}}
\newcommand{\W}{\mbox{$W~$}}

\newcommand{\xB}{\mbox{$x~$}}  

\newcommand{\Qsq}{\mbox{$Q^2~$}}
\newcommand{\et}{\mbox{$E_T~$}}
\newcommand{\kt}{\mbox{$k_T~$}}

\newcommand{\as}{\mbox{$\alpha_s~$}}
\newcommand{\ycut}{\mbox{$y_{\rm cut}~$}}

\newcommand{\GeV}{\mbox{\rm ~GeV~}}
\newcommand{\GeVx}{\rm GeV}

\newcommand{\GeVsq}{\mbox{${\rm ~GeV}^2~$}}

\newcommand{\mmm}{\mbox{$\cdot 10^{-3}$}}

\newcommand{\epem}{\mbox{$e^+e^-$}}
\newcommand{\ep}{\mbox{$ep~$}}
\begin{document}
\begin{titlepage}

%
\noindent
{\tt DESY 95-108    \hfill    ISSN 0418-9833} \\
\tt hep-ex/9506012 \\
{\tt June 1995}                  \\

\begin{center}


\vspace*{2cm}

\begin{Large}

{\bf  Transverse Energy and Forward Jet Production in the
      Low $x$ Regime at HERA
   }\\[1.5cm]
%
\vspace*{2.cm}
H1 Collaboration \\
\end{Large}

\vspace*{1cm}

\end{center}

\vspace*{1cm}

\begin{abstract}

\noindent
The production of transverse energy in deep inelastic scattering
is measured
as a function of the kinematic variables
\xB and \Qsq using the H1 detector at the \ep collider HERA.
The results are compared to the different predictions
based upon two alternative QCD evolution equations,
namely the
Dokshitzer-Gribov-Lipatov-Altarelli-Parisi (DGLAP) and the
Balitsky-Fadin-Kuraev-Lipatov (BFKL)
equations.
In a pseudorapidity interval which is central in the hadronic
centre of mass system
between the current and the proton remnant fragmentation region the
produced transverse energy increases with decreasing \xB for constant
\Qsq. Such a behaviour can be explained with a QCD calculation
based upon the BFKL ansatz.
The rate of forward jets, proposed as a signature
for BFKL dynamics, has been measured.

\vspace{1cm}
\centering{to be submitted to Phys. Lett.}

\end{abstract}
\end{titlepage}

\begin{Large} \begin{center} H1 Collaboration \end{center} \end{Large}
\begin{flushleft}
 S.~Aid$^{13}$,                   
 V.~Andreev$^{25}$,               
 B.~Andrieu$^{28}$,               
 R.-D.~Appuhn$^{11}$,             
 M.~Arpagaus$^{36}$,              
 A.~Babaev$^{24}$,                
 J.~B\"ahr$^{35}$,                
 J.~B\'an$^{17}$,                 
 Y.~Ban$^{27}$,                   
 P.~Baranov$^{25}$
% ...

\end{flushleft}
\begin{flushleft} {\it
% ...
$ ^{11}$ DESY, Hamburg, Germany$^a$ \\
% ...
$ ^{13}$ II. Institut f\"ur Experimentalphysik, Universit\"at Hamburg,
                                     Hamburg, Germany$^ a$  \\
% ...
$ ^{17}$ Institute of Experimental Physics, Slovak Academy of
                Sciences, Ko\v{s}ice, Slovak Republic$^ f$\\
% ...
$ ^{24}$ Institute for Theoretical and Experimental Physics,
                                                 Moscow, Russia \\
 $ ^{25}$ Lebedev Physical Institute, Moscow, Russia$^ f$ \\
% ...
$ ^{27}$ LAL, Universit\'{e} de Paris-Sud, IN2P3-CNRS,
                            Orsay, France\\
 $ ^{28}$ LPNHE, Ecole Polytechnique, IN2P3-CNRS,
                             Palaiseau, France \\
% ...
$ ^{35}$ DESY, Institut f\"ur Hochenergiephysik,
                              Zeuthen, Germany$^ a$\\
 $ ^{36}$ Institut f\"ur Teilchenphysik,
          ETH, Z\"urich, Switzerland$^ i$\\
% ...
$ ^a$ Supported by the Bundesministerium f\"ur
                                  Forschung und Technologie, FRG
 under contract numbers 6AC17P, 6AC47P, 6DO57I, 6HH17P, 6HH27I, 6HD17I,
 6HD27I, 6KI17P, 6MP17I, and 6WT87P \\
% ...
$ ^f$ Supported by the Deutsche Forschungsgemeinschaft\\
% ...
$ ^i$ Supported by the Swiss National Science Foundation\\
   } \end{flushleft}
\newpage

\section{Introduction}

The
electron-proton collider HERA has opened new kinematical regions
in the study of
Deep Inelastic Scattering (DIS): the regions of large four-momentum
transfer \Qsq (up to $\Qsq \approx 10^4 \GeVsq$) and  small Bjorken-\xB
(down to \xB $ \approx 10^{-4}$).
It has been suggested that the small \xB region may
be sensitive to new dynamic features
of QCD, e.g.~\cite{levin}.
The ZEUS and
H1 collaborations have observed \cite{z93f2,h1f2} that
the proton structure function $F_2$
exhibits a strong rise towards small Bjorken-$x$.
This rise has caused much debate on whether the HERA data are still in
a regime
where the QCD evolution of the parton densities can be described by the
DGLAP
(Dokshitzer-Gribov-Lipatov-Altarelli-Parisi)~\cite{dglap}
 evolution equations, or whether they extend into a new regime
where the QCD dynamics is
described by the BFKL
(Balitsky-Fadin-Kuraev-Lipatov)~\cite{bfkl}
evolution equation.
The BFKL evolution equation is expected to become applicable
in the small $x$ region, since it resums all leading
$\alpha_s\ln1/x$ terms in the perturbative expansion,
in contrast to the DGLAP equation.
Present $F_2$ measurements
do not yet
allow to discriminate between BFKL and conventional DGLAP
dynamics~\cite{akms,h1gluon}, and are perhaps
too inclusive a measure to be
a sensitive discriminator.
Hadronic final states may give additional information and
could be more sensitive to the parton evolution~\cite{webber,mueller,dhotref}.
In this paper we use the H1 detector to
study properties of the hadronic final state, namely the transverse energy
flow and the production of forward jets,
to test
the BFKL versus DGLAP hypothesis.
Throughout this paper,
``forward'' refers to the region around the proton direction in the
HERA laboratory frame, and
transverse quantities are calculated with respect to the
proton direction.

\begin{figure}[htb] \centering \unitlength 1mm
\epsfig{file=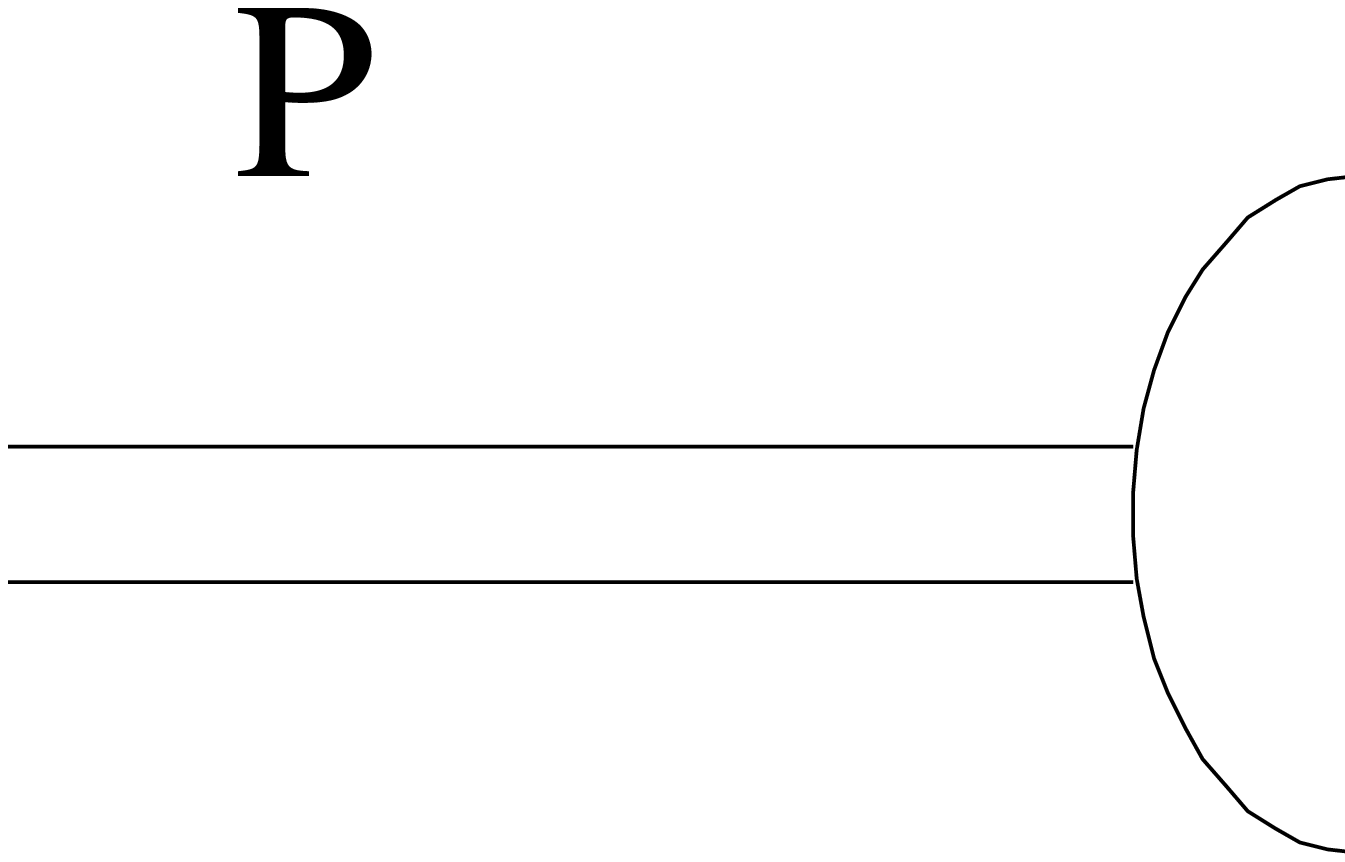,bbllx=0pt,bblly=0pt,bburx=2000pt,bbury=2000pt,height=10cm}
\vspace{-1.5cm}
\scaption{Parton evolution in the ladder approximation. The
selection of forward jets in DIS events is illustrated.}
\label{HOTSPOT}
\end{figure}

For events at low $x$, hadron production in the
region between the current jet and the
proton remnant is expected to be
sensitive to the effects of the BFKL or DGLAP dynamics.
At lowest order the BFKL and DGLAP  evolution equations effectively resum the
leading logarithmic $\alpha_s\ln1/x$ or
$\alpha_s\ln Q^2$ contributions respectively.
In an axial gauge this
amounts to a resummation of ladder diagrams of the type shown in
Fig.~\ref{HOTSPOT}.
This shows that before a
quark is struck by the virtual photon,
a cascade of partons may be emitted.
The fraction of the proton momentum carried by the emitted partons,
$x_i$, and their transverse momenta,
$k_{Ti}$,
are indicated in the figure.
In the leading log DGLAP scheme
this parton cascade follows a strong ordering in transverse
momentum
$k_{Tn}^2 \gg k_{Tn-1}^2 \gg... \gg k_{T1}^2$,
while there is only a soft
(kinematical) ordering for the fractional momentum $x_n<x_{n-1}<...<x_1$.
In the BFKL scheme the cascade follows a strong ordering in fractional
momentum
$x_n \ll x_{n-1} \ll... \ll x_1$,
while there is no ordering in transverse
momentum\cite{muellercarg}.
 The transverse momentum follows a kind of random walk
in $k_{T}$ space:
the value of $k_{Ti}$ is close to that of $k_{Ti-1}$, but it
can be both larger or smaller~\cite{bartels}.
As a consequence, BFKL evolution is expected to produce more
transverse energy $E_T$
than DGLAP
evolution~\cite{durham1,durham2}
in the region between the struck quark
and the remnant for low \xB events.
Intriguingly, a numerical
BFKL calculation of the transverse energy flow
\cite{durham2}
came out close to the measurements published by H1 \cite{h1flow2},
while the DGLAP prediction is too low.
In addition, the DGLAP calculation
predicts an increase in transverse energy with rising $x$,
while the BFKL calculation predicts the opposite \cite{durham1}.
With increased statistics,
the $x$ dependence of
this effect can be studied in more detail in this paper.

Another possible signature of the BFKL dynamics is the rate of jets
with transverse momentum
$\kjet \approx Q$ and the momentum fraction of the jet,
$\xjet = \Ejet/E_p$, large compared with Bjorken-$x$
\cite{mueller,dhotref,hotref}.
Here $\Ejet$ and $E_p$ are the energies of the jet and the incoming
proton respectively.
Due to the strong ordering in DGLAP evolution,
the condition $\kjet^2\approx Q^2$ suppresses the phase space
for jet production.
However jet production from BFKL evolution is governed by
the ratio $\xjet / x$, which is large.
Hence the rate of events with a jet satisfying the selection is predicted to
be higher for the BFKL than for the DGLAP scenario.


Apart from numerical calculations,
predictions for final state observables are also available
as Monte Carlo models, based upon QCD phenomenology.
In this report we consider two of the currently available
Monte Carlo programs:
the MEPS (Matrix Elements plus Parton Showers) and CDM (Colour Dipole Model)
models.
Both provide good descriptions of
fixed target DIS
and \epem data \cite{perform}.
The CDM model~\cite{ariadne} provides an implementation of the colour dipole
model of a chain of independently radiating dipoles formed by
emitted gluons~\cite{dipole}.
Photon-gluon fusion events are not described by this picture
and are added at a rate given
by the QCD matrix elements~\cite{lepto}.
The CDM description
of gluon emission is similar to that of the BFKL evolution,
because the gluons emitted by the dipoles
do not obey strong ordering in \kt~\cite{bfklcdm}.
The CDM does not explicitly
make use of the BFKL evolution equation, however.
The MEPS model is an option
of the LEPTO generator~\cite{lepto} based on DGLAP dynamics.
MEPS incorporates the QCD matrix elements up to first order, with additional
soft emissions generated by adding leading log parton showers.
The emitted partons are strongly ordered in $k_T$.
Both Monte Carlo programs
use the Lund string model \cite{string} for hadronizing the
partonic final state.
The
parton density parametrization used here is that of
MRSH \cite{mrsh}, which results in a good description of the
HERA $F_2$ measurements\cite{z93f2,h1f2}.

\section{Detector Description}

A detailed description of the H1 apparatus
can be found elsewhere~\cite{h1nim}.
The following briefly describes the components of the detector relevant
to this
analysis.

The hadronic energy flow and the scattered electrons are measured with a
liquid argon~(LAr) calorimeter and a
backward electromagnetic lead-scintillator calorimeter (BEMC).
The LAr calorimeter~\cite{larc}  extends over the polar angular range
$4^\circ < \theta <  153^\circ$ with full azimuthal coverage, where
 $\theta$ is defined with respect to the proton
beam direction ($+z$ axis).
It consists of an electromagnetic section with lead absorbers
and a hadronic section with steel absorbers.
Both sections are highly segmented in the transverse and
longitudinal directions
with about 44000 cells in total.
The total depth of both calorimeters
varies between 4.5 and 8 interaction lengths for $\theta <  125^\circ$.
Test beam measurements of the LAr~calorimeter modules show an
energy resolution
of $\sigma_{E}/E\approx 0.50/\sqrt{E\;[\GeVx]} \oplus 0.02$  for
charged pions~\mbox{\cite{h1pi}}.
The hadronic energy measurement is performed
by applying a weighting technique in
order to account
for the non-compensating nature of the calorimeter.
The
absolute scale of the hadronic energy measurement
is presently known to $5\%$, as
determined from studies of the
transverse momentum ($p_T$) balance in DIS events.

The BEMC (depth of 22.5 radiation lengths or 1
interaction length) covers the backward region of the detector,
$151^\circ < \theta < 176^\circ$.
The BEMC energy scale for electrons is known to an accuracy of $1.7\%$.
Its resolution is given by
$\sigma_{E}/E = 0.10/\sqrt{E\;[\GeVx]} \oplus 0.42/E[\GeVx] \oplus 0.03$
\cite{f2pap}.

The calorimeters are
surrounded by a superconducting solenoid which provides a uniform
magnetic field of $1.15$ T parallel to the beam axis in the tracking region.
Charged particle
tracks are measured in a central drift chamber and the forward
tracking system,
covering the polar angular range $ 7^\circ < \theta < 165^\circ$.
A backward proportional chamber (BPC), in front of the BEMC with an angular
acceptance of $155.5^\circ < \theta < 174.5^\circ$ serves to identify electrons
and to precisely measure their direction.
Using information from the BPC, the BEMC and the reconstructed event vertex the
polar angle of the scattered electron is known to a precision of 2 mrad.

\section{
Event Selection and Kinematics}

The data used in this analysis were collected in 1993,
with electrons of energy $E_e=26.7\GeV$ colliding with
protons of energy $E_p=820$~GeV, resulting in a total centre
of mass energy of $\sqrt{s}=296 \GeV$.
The data correspond to an integrated luminosity of 320 nb$^{-1}$.
For this analysis DIS events with
$Q^2 < 100$~GeV$^2$ are used, in which the scattered electron is observed in
the BEMC.
The events are triggered by requiring a cluster of more than 4~GeV
in the BEMC.
After reconstruction, DIS events are selected in
the following way:

\begin{itemize}
\item The scattered electron, defined as the most energetic BEMC cluster,
  must have an energy $E'_e$ larger than 12~GeV and a polar angle $\theta_e$
  below $173^\circ$ in order to ensure high trigger efficiency and a small
  photoproduction background~\cite{h1f2}.

\item The lateral size of the electron
  cluster, calculated as the energy weighted radial distance of the cells from
  the cluster centre, has to be smaller than 4~cm.
  The cluster must be associated
  with a reconstructed BPC space point
  which must lie within 4~cm of the cluster centre of gravity.
  Further reduction of photoproduction background
  and the removal of events in which an energetic photon is radiated off
  the incoming electron
  is achieved by
  requiring $\sum_j{(E_j-p_{z,j})} >30\GeV$ \cite{h1f2},
  with
  the sum extending over all particles $j$ (measured calorimetrically)
  of the event.

\item
  The radial coordinate of the BPC hit must be less than
  60~cm, corresponding
  to an electron angle above $157^\circ$ with respect to the nominal
  interaction point, ensuring full containment of the electron shower
  in the BEMC.

\item
  The $z$ position of the
  event vertex reconstructed from charged tracks has to be within 30~cm
  of the average of all collision events.

\item The energy in the forward region
  ($4.4^\circ < \theta < 15^\circ$) has to be larger than 0.5~GeV in order
  to exclude diffractive-like events with large rapidity gaps in the forward
  region \cite{gap,h1flow2}.

\end{itemize}

The kinematic variables are determined using
information from the scattered electron:
$ Q^2 = 4\,E_e \, E'_e\cos^2(\theta_{e}/2)$ and
$ y = 1-(E'_e/E_e)\cdot \sin^2(\theta_{e}/2)$.
The scaling variable $x$
is then derived via $x=Q^2/(ys)$, and
the hadronic invariant mass squared is
$W^2=m_p^2+sy-Q^2$.

\begin{itemize}
\item
As the
precision of the $y$ measurement degrades with $1/y$,
a cut $y > 0.05$ is imposed.
Events suffering from QED radiation or from a
badly reconstructed electron
are removed by requiring that they also fulfil
this cut if $y$ is calculated from the measured hadrons.
\end{itemize}



\section{Transverse Energy Flows}

The event sample for the energy flow measurements,
in which 60\% of the total luminosity has been used,
contains 9529 DIS events with
$5\GeVsq < Q^2 < 100$~GeV$^2$ and with $10^{-4} < x < 10^{-2}$.
The measurements are performed in
the hadronic centre of mass system (CMS),
that is
the rest system of the proton and the exchanged
boson. The orientation of the CMS is such that the direction of
the exchanged boson defines the positive $z^\prime$ axis.

\begin{figure}[p]
   \centering
   \epsfig{file=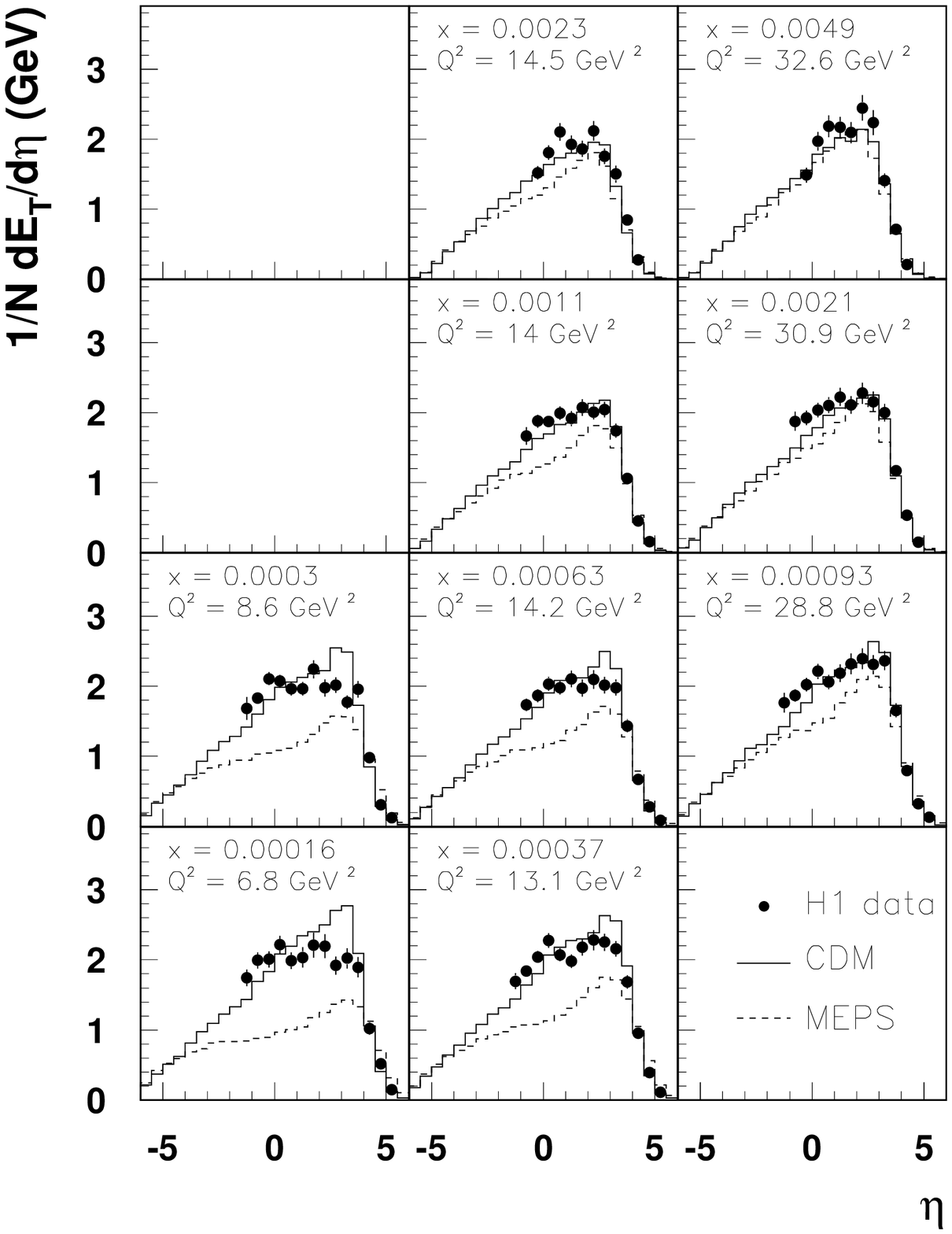,width=16cm,%
    bbllx=63pt,bblly=120pt,bburx=524pt,bbury=718pt,clip=}
   \scaption{
            The flow of transverse energy \et in the CMS as a function
            of pseudorapidity $\eta$.
            The remnant direction is to the left.
            Shown are data for nine different kinematic bins
            (see Table~1).
            The distributions are normalized to the number of events
            $N$ in each case. Only statistical errors are shown.
            For comparison, the models CDM (full line) and MEPS (dashed)
            are overlayed.}
   \label{etcms}
\end{figure}

The \xB -- \Qsq plane is divided into three slices with
constant \Qsq and varying $x$,
resulting in 9 kinematic bins (see Table~\ref{tab:et}).
In Fig.~\ref{etcms}, the flow of transverse energy,
as measured with the LAr and BEMC calorimeters, is shown
as a function of pseudorapidity
\mbox{$\eta = -\ln\tan (\theta /2)$}
for the individual kinematic bins.
The data are corrected for
detector effects
using the CDM generator and a full simulation
of the H1 detector response.
The data in general exhibit a moderate slope,
extending from the current region towards the proton remnant up
to the edge of the detector acceptance.
For large \xB and \Qsq the data are reasonably well
described by the two models under consideration, MEPS
and CDM.
At small \xB and \Qsq however, the models, in particular MEPS,
deviate significantly
from the data.
The MEPS model
generates
too little $E_T$.
The CDM
is able to describe the \et flow towards the remnant, but
overestimates the \et in the current region, around $\eta=3$.
We note that in events where the hard subprocess leads to
two hard jets in the detector, both models provide a
good description of the energy flow \cite{xg} around the jets.

\begin{table}
\begin{center}
\begin{tabular}{|c|c|c|c|c|c|}
  \hline
\xB $/10^{-3}$ & $\Qsq / \GeVsq$ & \av{\xB} $/10^{-3}$ & $\av{\Qsq}/\GeVsq$ &
\multicolumn{2}{c|}{\av{\et} (GeV/unit rapidity)} \\
\hline
         &      &       &        & measured               & BFKL calc. \\
  \hline
 0.1--0.2 & 5--10 & 0.16  &  6.8   & $ 2.12 \pm 0.15 $ & 1.82\\
 0.2--0.5 & 6--10 & 0.30  &  8.6   & $ 2.09 \pm 0.12 $ & 1.69\\
\hline
 0.2--0.5 & 10--20& 0.37  & 13.1   & $ 2.17 \pm 0.12 $ & 1.77\\
 0.5--0.8 & 10--20& 0.63  & 14.2   & $ 1.95 \pm 0.12 $ & 1.68\\
 0.8--1.5 & 10--20& 1.1   & 14.0   & $ 1.87 \pm 0.11 $ & 1.57\\
 1.5--4.0 & 10--20& 2.3   & 14.5   & $ 1.67 \pm 0.12 $ & \\
\hline
 0.5--1.4 & 20--50& 0.93  & 28.8   & $ 2.12 \pm 0.12 $ & \\
 1.4--3.0 & 20--50& 2.1   & 30.9   & $ 1.99 \pm 0.13 $ & \\
 3.0--10  & 20--50& 4.9   & 32.6   & $ 1.75 \pm 0.17 $ & \\
  \hline
\end{tabular}
\end{center}
\caption[]{
{\footnotesize
Values of \av{\et} measured in the CMS pseudorapidity
interval from $-0.5$ to $0.5$ as a function of \xB and \Qsq.
The errors quoted contain the statistical and the
systematic point-to-point errors added in quadrature.
An overall uncertainty of 9\% is not included.
Also given are perturbative BFKL predictions calculated
according to \cite{durham2} (see text).
}
}
\label{tab:et}
\end{table}

\begin{figure}[htb]
   \centering
   \begin{picture}(0,0) \put(0,0){{\bf a)}} \end{picture}
   \epsfig{file=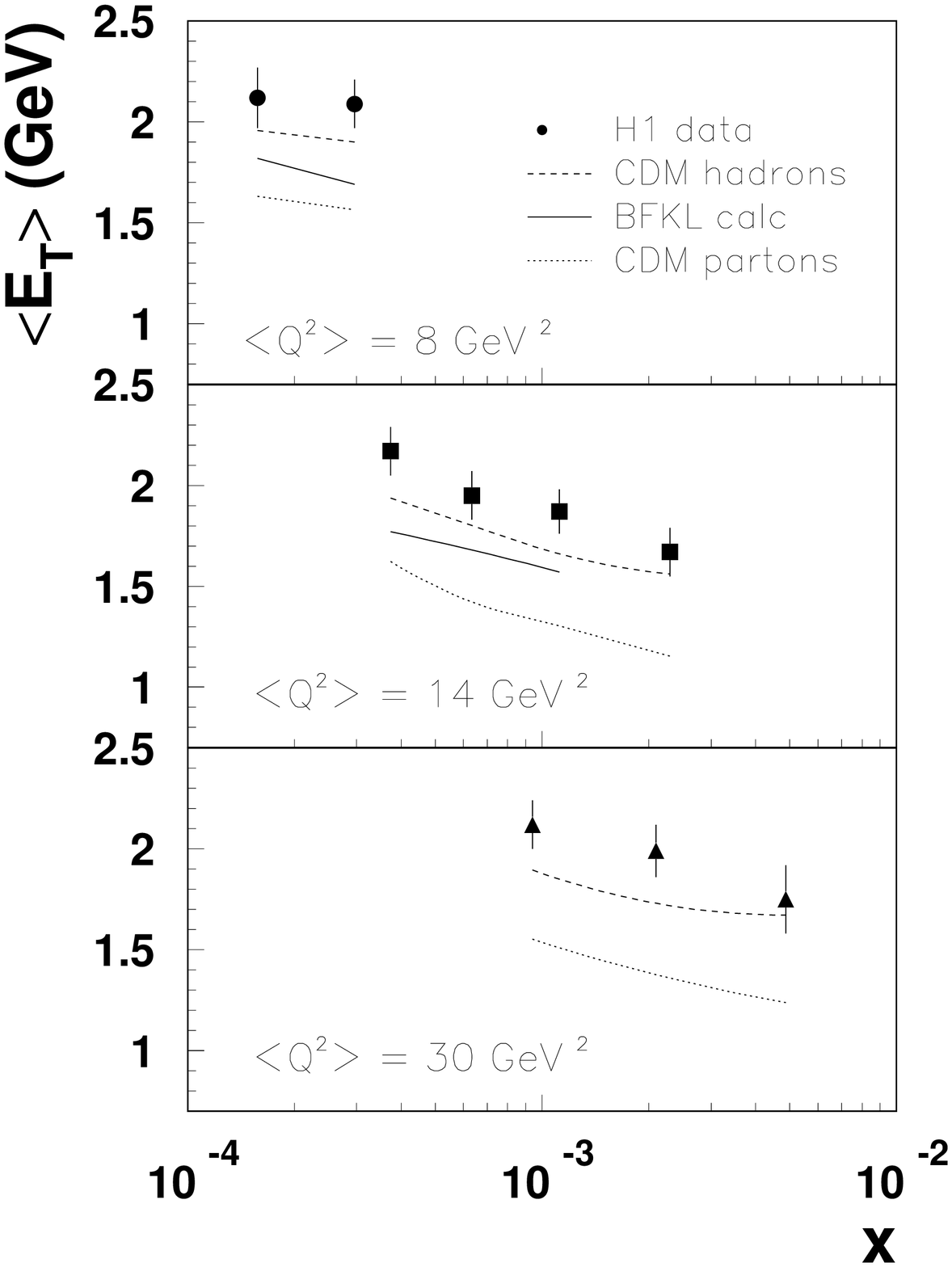,width=7.5cm,%
    bbllx=10pt,bblly=75pt,bburx=551pt,bbury=765pt,clip=}
   \begin{picture}(0,0) \put(0,0){{\bf b)}} \end{picture}
   \epsfig{file=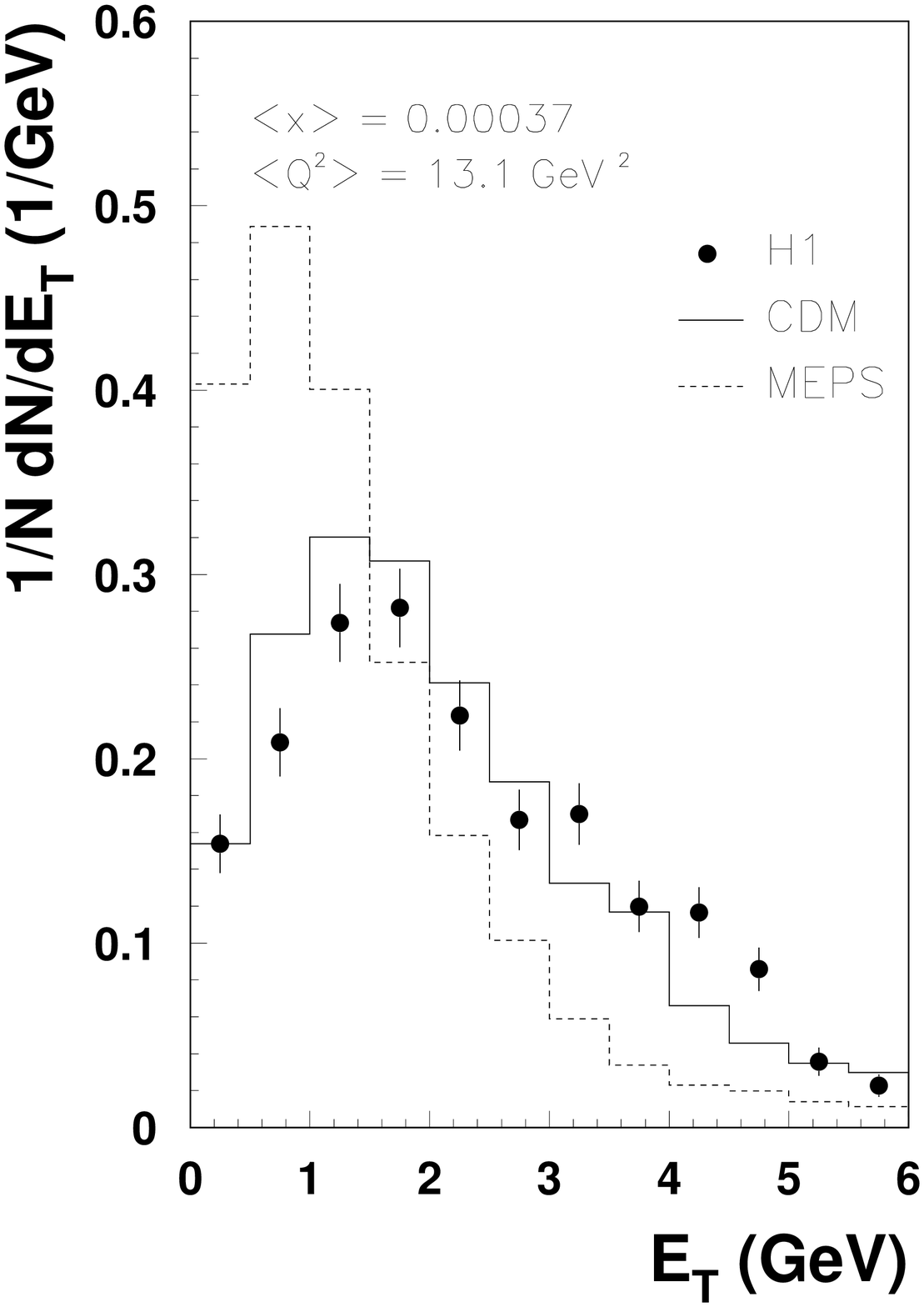,width=7.5cm,%
    bbllx=10pt,bblly=75pt,bburx=551pt,bbury=765pt,clip=}
   \scaption{
            {\bf a)}
            Transverse energy as a function of \xB for three
            different values of \Qsq.
            The transverse energy \av{\et} is measured in the CMS
            in the pseudorapidity range from $-0.5$ to $+0.5$.
            Shown are the H1 data, corrected for detector effects,
            the BFKL calculation and
            the prediction from the CDM before (partons) and
            after (hadrons) hadronization.
            The error bars contain the statistical and the systematic
            point-to-point errors added in quadrature. An overall
            scale uncertainty of 9\% is not shown.
            {\bf b)}
            The uncorrected
            distribution of the \et (as defined above) for the
            events of the kinematic bin with
            $\av{\xB}=0.37\mmm$ and $\av{\Qsq}=13.1 \GeVsq$.
            The data are compared with the CDM and
            MEPS models, including a full simulation of the H1 detector.
             }
   \label{etx}
\end{figure}

In order to quantify the evolution of \et with \xB and $Q^2$,
the average \et per unit of pseudorapidity
is measured in the central pseudorapidity range
$-0.5 < \eta < 0.5$.
Note that this corresponds roughly to the forward
range $2<\eta_{\rm lab}<3$ in the laboratory frame.
The measured values of $\average{\et}$ as a function
of \xB are shown for the three \Qsq slices in Fig.~\ref{etx}a,
and summarized in Table~\ref{tab:et}.
They are of the order of 2 GeV per unit of pseudorapidity.
For $\average{\Qsq}\approx 14 \GeVsq$,
the $\average{\et}$ drops by about
25\%
when going from $\av{\xB}=0.37 \mmm$ to $2.3 \mmm$.
Though less significant,
the data at $\av{\Qsq}\approx 8 \GeVsq$
and at $\av{\Qsq}\approx 30 \GeVsq$ confirm
these $\average{\et}$ measurements.
For fixed $x$, the level of \et rises slightly with \Qsq
\footnote{When \W and \Qsq are chosen as kinematic variables,
$\av{\et}$ is found to rise with \W and to be almost independent of \Qsq.}.

The errors shown in Fig.~\ref{etx}a
are the statistical and
systematic point-to-point errors added in quadrature, which
are typically
5~\% each.
The corrections for detector effects
are typically 20\%.
The uncertainty in the BEMC calibration (1.7\%) leads to a
sizeable effect (4\%) in the bin of largest \xB and $Q^2$.
Small
effects from QED radiation (typically 1--2\%)
and from
photoproduction background
in the lowest \xB and \Qsq bin (3\%),
estimated from Monte Carlo simulations,
are not corrected for but absorbed
in the systematic error.
The model dependence of the correction was investigated
with the CDM and MEPS models and leads to a 4\%
point-to-point error and a 5\% overall error.
In addition, an overall scale error of 7\% arises from
the calorimeter calibration (5\%), and from varying
details of the analysis method, such as the treatment of
noise in the calorimeter,
clustering of calorimeter cells,
and the simulation of
the calorimeter response,
affecting the result by 5\% in the forward region.

Fig.~\ref{etx}a and Table~\ref{tab:et}
display the prediction for $\av{\et}$ from
a BFKL based QCD calculation \cite{durham2}, yielding values
of the order of 1.7 GeV per unit of pseudorapidity and
with similar slopes as a function of \xB as seen in the data.
The validity of the calculation \cite{durham2} is
limited to the range shown.
A DGLAP based calculation yields an $\av{\et}$ of
around 0.4 GeV \cite{durham2} with the opposite $x$ dependence.
These calculations however
do not include the non-perturbative hadronization.
This
can presently
only be included using Monte Carlo models which contain both
the perturbative QCD evolution and a phenomenological
hadronization model.
Since the CDM provides a reasonable description of
the data, and since the perturbative BFKL
calculation of $\av{\et}$ is close
to the partonic final state $\av{\et}$
produced perturbatively in the CDM
(see Fig.~\ref{etx}a), the CDM may provide an estimate of the effect
of hadronization on the BFKL $\av{\et}$ calculation.
That effect can be seen in
Fig.~\ref{etx}a, where the $\av{\et}$ according to the CDM is
plotted before and after hadronization.
It amounts to an increase in $\av{\et}$ of
0.3-0.4 GeV and is rather independent of $x$.
When taking into account the hadronization effect as
modelled by the CDM, the BFKL calculation is in good
agreement with the data.
In this context it is
interesting to observe
that the Monte Carlo model without \kt
ordering, the CDM,
is consistent with the BFKL
calculation for the perturbative part,
and that it
is able to describe the measured
magnitude and \xB dependence of $\av{\et}$
reasonably well.

It has to be stressed that the model predictions
depend on a variety of parameters which can be tuned \cite{h1flow2}.
For this analysis the model settings were chosen such as to render
a good description of the
measured energy flow at large \xB and $Q^2$,
a kinematic region where theoretical uncertainties are
minimal.
CDM and MEPS were used in the program implementations
ARIADNE 4.03 and LEPTO 6.1, respectively.
In the MEPS model,
divergences of the matrix element are avoided
with a cut-off for parton-parton invariant masses,
$m_{ij} > \ycut \cdot \W$.
For this analysis the MEPS cut-off was parametrized such
as to follow
the limit
at which the order \as contribution
exceeds the total cross section
within a margin of 2 \GeV.
For the CDM
the standard value of the parameter \ycut
($\ycut=0.015$) is used to
regulate the admixture
of boson-gluon fusion events according to the matrix element.
Turning off the boson-gluon fusion admixture, or
using the same cut-off prescription as described above
has little influence on the CDM predictions.


Improvements to the MEPS model are conceivable, for example
with an improved scheme for matching parton showers and
matrix elements, by changing the arrangement of color connections,
or by changes to the remnant fragmentation which is not well
tested at small $x$. Whether or not such improvements can
result in an acceptable description of the energy flows while
maintaining a good description of other final
state observables can presently not be judged.
Therefore the failure of the MEPS model to describe $\av{\et}$
cannot be
unambiguously identified with the fact that its parton shower
evolution is based upon the DGLAP evolution.
For the same reasons the success of the CDM in this respect,
where the parton cascade does not obey strong \kt
ordering as in the BFKL evolution, is intriguing but
may be fortuitous.

Having identified
the average $E_T$ as a sensitive observable to test
QCD predictions,
it is
interesting as well to study the shape of the
\et distribution.
The
distribution of the observed \et in the central pseudorapidity
bin  $-0.5 <\eta<0.5$ is shown in Fig.~\ref{etx}b for
a kinematic bin at low \xB and $Q^2$.
The CDM gives a reasonable description of the data,
and the MEPS model produces events accumulating
at lower values of \et than the majority of the data events.
The difference in \av{\et} between the data and the MEPS model
does not seem to stem from a tail in the data.


\section{Forward Jets}

\begin{figure}[p]
\centering
   \begin{picture}(0,0) \put(0,0){{\bf a)}} \end{picture}
   \begin{picture}(0,0) \put(300,150){{\Large electron}} \end{picture}
   \begin{picture}(0,0) \put(-50,100){{\Large forward}} \end{picture}
   \begin{picture}(0,0) \put(-50,80){{\Large jet}} \end{picture}
   \begin{picture}(0,0) \put(100,30){{\Large liquid argon}} \end{picture}
   \begin{picture}(0,0) \put(270,80){{\Large BEMC}} \end{picture}
   \epsfig{figure=event.landscape.eps,height=8cm,angle=90,%
    bbllx=48pt,bblly=745pt,bburx=393pt,bbury=283pt,clip=/}
   \epsfig{figure=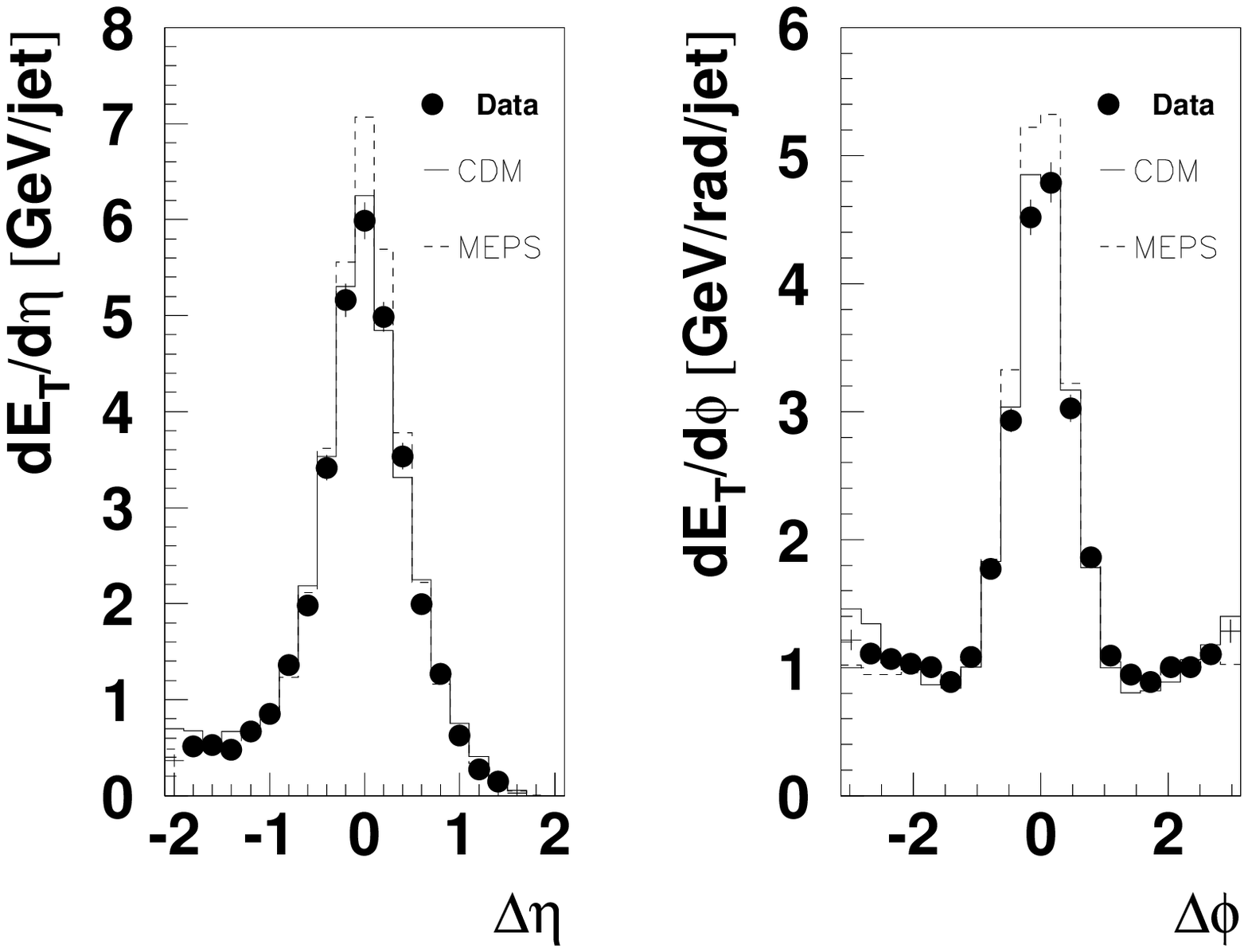,height=8cm}
   \begin{picture}(0,0) \put(-300,0){{\bf b)}} \end{picture}
   \begin{picture}(0,0) \put(-150,0){{\bf c)}} \end{picture}
\scaption{
   {\bf a)} DIS event event with a forward jet in the H1 detector.
The protons are incident from the right, electrons from the left.
The scattered electron is detected
in the backward electromagnetic calorimeter (BEMC)
with an angle of $166^0$
and an energy of 18.9 GeV.
The forward jet is observed in the liquid argon
calorimeter and has an angle $\thjet = 11^0 $
and an energy $\Ejet = 65 $ GeV.
Averaged over all events with a selected forward jet, the
transverse energy flow around the forward jet axis is shown in
{\bf b)} as a
function of $\Delta\eta$, integrated over $|\Delta \phi| < 1.0$
and in
{\bf c)}
as a function of $\Delta\phi$, integrated over $|\Delta \eta| < 1.0$.
Here  $\Delta\eta$ and $\Delta \phi$ are measured with respect
to the reconstructed jet axis.}
 \label{HOTJETS}
\end{figure}

We have studied DIS events at small \xB
which have a jet
with large $\xjet$ \cite{joachim}.
A cone algorithm is used to find jets, requiring an $E_T$ larger than
5~GeV in a cone of radius
$ R = \sqrt{\Delta\eta^2 + \Delta\phi^2} = 1.0$ in the space of
 pseudo-rapidity $\eta$ and azimuthal angle $\phi$
in the HERA frame of reference.
In addition to the selections given in
section 3, the requirement $y > 0.1$ was imposed
to ensure that the jet of the
struck quark is well within the central region of the detector
and is expected to have a jet angle larger than $ 60^0$.
In this  sample of DIS events with $Q^2 \approx 20~{\rm GeV}^2$ and
$2\!\cdot\!10^{-4}\!<\!x\!<\!2\!\cdot\!10^{-3}$ we have counted
events which have a ``forward'' jet defined by
$\xjet\!>\!0.025$, $0.5\!<\! \pjet^2/\Qsq\!<\!4$,
$6^0 < \thjet< 20^0$ and $\pjet> 5 $ GeV,
where $\pjet$ is the transverse momentum of the jet.
A typical event with a high energy forward jet is shown in
Fig.~\ref{HOTJETS}a.
The transverse energy flow around the forward jet axis,
averaged over all selected events,
is shown versus $\eta$ and $\phi$ in Figs.~\ref{HOTJETS}b and
\ref{HOTJETS}c.
Distinct jet profiles are observed, which are
well described by the Monte Carlo models.

\begin{table}[t]
\centering
\begin{tabular}{|c|c|c|c|c|} \hline
$x$ range & data & MEPS &CDM & $\sigma(ep\rightarrow {\rm jet}+X)$ \\
        &  events    & events & events & (pb) \\ \hline
$2\cdot 10^{-4} - 1\cdot 10^{-3}$ &$271$ & 141 & 282 &
$709\pm42\pm166$ \\ \hline
$1\cdot 10^{-3} - 2\cdot 10^{-3}$ &$ 158 $ & 101 & 108 &
$475 \pm 39 \pm 110$ \\ \hline
\end{tabular}
\caption[]
{\label{HOTTABLE}
{\footnotesize
Numbers of observed DIS events with a selected forward jet,
corrected for radiative events faking this signature.
These may be directly compared with the
expectations from the Monte Carlo models.
The measured cross section $ep\rightarrow {\rm jet} +X$
for  forward jets
is also given.
The errors reflect the statistical and systematic uncertainties.}}
\end{table}

The resulting number of
events observed with at least one forward jet in the kinematical
region  $160^0<\theta_e<173^0$ and $E_e>12$ GeV
is given in Table~\ref{HOTTABLE} and compared to
expectations of the MEPS and CDM models after detector simulation.
The data are corrected for photoproduction background and
radiative events, which due to the changed kinematics at
the hadron vertex can eject a jet in the forward direction.
About 4\% of the data events were
found to contain two forward jets.
Results on jet production for both models have been compared with data
in previous analyses \cite{h1alphas,h1disjet}. In particular the MEPS model
was found to show  good agreement with the data outside the
forward region.
In the kinematic range studied here
the CDM generally describes the data better than the MEPS model.
The predictions were found not to
depend significantly on the uncertainties
of the proton structure function.
However,
increasing the $\xjet$ cut from 0.025 to 0.05 reduces the total number of
events with forward jets to 46 for CDM, to 77 for MEPS
and to 105 for the data,
hence CDM does not describe the rate of high energy jets.

The measured cross section for forward jets
satisfying the cuts given above is also presented in
Table~\ref{HOTTABLE}.
It has been corrected for detector effects using the CDM.
The systematic errors include effects from
DIS event selection,
the calorimeter energy scale (5\%), the jet angle bias (10 mrad),
the proton structure function,
and a global normalization uncertainty of 4.5\%.
Event pile-up effects were found to be negligible.
The systematic errors on the two data points are largely correlated.
The ratio of the jet cross section for the low $x$ to the  high $x$ bin
is $1.49\pm 0.25$.

The precision of the data does not yet allow  a firm
conclusion to be drawn.
We note, however, that the
forward jet cross section is larger
in the low $x$ bin than in the high $x$ bin.
This is  expected from BFKL dynamics
as an analytical calculation~\cite{dhotref,deduca}
at the parton level demonstrates:
in the kinematical region selected the ratio of the cross
sections in the low $x$ bin to high
$x$ bin is 1.62 for a calculation including BFKL evolution,
compared to 1.03 for a calculation without gluon
emission from the ladder in Fig.~\ref{HOTSPOT}.

\section{Conclusions}

In order to shed light on the QCD mechanism
responsible for
parton evolution in the regime of small Bjorken-$x$,
the production of particles and of jets in the forward region
has been measured at HERA.
A forward jet selection designed to enhance the yield in the case
of BFKL evolution, and to suppress the yield for DGLAP evolution,
results in a rate of observed forward jets
compatible with the BFKL expectation.
A firm conclusion
on the growth with \xB  however
would necessitate a larger data sample.

The flow of transverse energy versus pseudorapity
in the hadronic centre of
mass system has been measured as a function
of the
kinematic variables $x$ and $Q^2$.
The observed magnitude and \xB dependence of the average
transverse energy is in agreement with a perturbative
QCD calculation based upon the BFKL mechanism,
assuming a non-perturbative contribution from hadronization
as predicted by the colour dipole model.
Though the BFKL mechanism provides a natural explanation
of the data, it is also possible that the currently used
hadronization scheme is inadequate, and that improved
DGLAP based models may also be able to describe the
observed features of transverse energy production.
The data, which are corrected for detector effects,
provide important constraints for further development
of the understanding of QCD at small $x$.

\bigskip\bigskip
\noindent
{\bf Acknowledgements.}
{\footnotesize We are grateful to the HERA machine group whose outstanding
efforts made this experiment possible. We appreciate the
immense effort of the engineers and technicians who
constructed and maintained the detector. We thank the funding
agencies for financial support. We acknowledge the support of the
DESY technical staff. We also wish to thank the DESY directorate
for the hospitality extended to the non-DESY members of the
collaboration.
We thank A.D. Martin and
P.J. Sutton for repeating the published BFKL calculations
for the kinematic bins covered by this analysis, and G. Ingelman
for helpful discussions on the Monte Carlo models.}

%
%

%
\end{document}